\documentclass{article}
\usepackage[utf8]{inputenc}
\usepackage[english]{babel}
\usepackage{amsmath}
\usepackage{color}
\usepackage{amsfonts}
\usepackage{amssymb}
\usepackage{graphicx}
\usepackage{listings}
\usepackage{hyperref}
\usepackage{adjustbox}
\usepackage{glossaries}
\usepackage{fancyvrb}
\usepackage[binary-units]{siunitx}
\usepackage{authblk}

\DeclareSIUnit{\belmilliwatt}{Bm}
\DeclareSIUnit{\dBm}{\deci\belmilliwatt}

\usepackage{subcaption}
\usepackage{fancyvrb}
\usepackage{eurosym}
\usepackage{tikz}
\usepackage{pgfplots}
\usetikzlibrary{calc}
\usepackage{datetime} 
\usepackage{chngcntr}

\definecolor{darkgreen}{RGB}{25,185,25}

\makeatletter
\newcommand{\verbatimfont}[1]{\def\verbatim@font{#1}}%
\makeatother

\date{}
\verbatimfont{\scshape}

\title{Electromagnetic fault injection against a System-on-Chip, toward new micro-architectural fault models}
\author[1]{Thomas Trouchkine}
\author[2]{Sébanjila Kevin Bukasa}
\author[2]{Mathieu Escouteloup}
\author[3]{Ronan Lashermes}
\author[1]{Guillaume Bouffard}

\affil[1]{ANSSI}
\affil[2]{INRIA/CIDRE}
\affil[3]{INRIA/SED\&LHS}

\newacronym{SoC}{SoC}{system-on-chip}
\newacronym{MMU}{MMU}{memory management unit}
\newacronym{OS}{OS}{operating system}
\newacronym{ALU}{ALU}{arithmetic logic unit}
\newacronym{CPU}{CPU}{central processing unit}
\newacronym{JTAG}{JTAG}{Joint Test Action Group}
\newacronym{PC}{PC}{program counter}
\newacronym{TLB}{TLB}{translation lookaside buffer}
\newacronym{SCU}{SCU}{snoop control unit}
\newacronym{ISA}{ISA}{instruction set architecture}
\newacronym{MAIR}{MAIR}{memory attribute indirection registers}
\newacronym{EMFI}{EMFI}{electromagnetic fault injection}
\newacronym{MAC}{MAC}{message authentication code}
\newacronym{IO}{IO}{input/output}
\newacronym{PTE}{PTE}{page table entry}
\newacronym{RPi3}{RPi3}{Raspberry Pi 3 B}
\newacronym{ROP}{ROP}{return-oriented programing}
\newacronym{PTD}{PTD}{page table directory}

\begin{document}
\maketitle

\begin{abstract}
  \Gls{EMFI} is a well known technique used to disturb the behaviour of a chip
  for weakening its security. These attacks are mostly done on simple
  microcontrollers. On these targets, the fault effects are relatively simple and
  understood.

  Exploiting \gls{EMFI} on modern \glspl{SoC}, the fast and complex chips
  ubiquitous today, requires to understand the impact of such faults. In this
  paper, we propose an experimental setup and a forensic process to create
  exploitable faults and assess their impact on the \gls{SoC} micro-architecture.

  On our targeted \gls{SoC} (a BCM2837), the observed behaviours are radically
  different to what were obtained with state-of-the-art fault injection attacks on
  microcontrollers. \gls{SoC} subsystems (L1 caches, L2 cache, \gls{MMU}) can be
  individually targeted leading to new fault models. We also highlight the
  differences in the fault impact with and without an \gls{OS}. This shows the
  importance of the software layers in the exploitation of a fault. By this
  work, we demonstrate that the complexity and the speed of \glspl{SoC} do not
  protect them against hardware attacks.

  To conclude our work, we introduce countermeasures to protect the \gls{SoC}
  caches and \gls{MMU} against \gls{EMFI} attacks based on the disclosed faults
  effect.

\end{abstract}

\glsresetall{}

\section{Introduction}

From a world where each application has its dedicated hardware support (high
performance chips for desktop computers, low power chips for mobile phones,
hardened chips for smartcards, \textit{etc.}), more and more applications
are today executed on a same device: the smartphone. They are powered by fast and
low-power components, the \glspl{SoC}, where a set of modules (modem, graphic
and sound card, Flash memory, \textit{etc.}) takes place on the same silicon
layout as the \gls{CPU}. With the democratization of smartphones, more sensitive
processes are done on the same component as the usual non sensitive processing. Aware of
this risk, vendors invest a lot of work into the software layer hardening
with numerous security mechanisms embedded into the \gls{OS} (such as Android or
iOS). However, the hardware layer misses the same attention: in particular it is
not protected against physical attacks, especially against fault injection
attacks.

Fault injection attacks is a well known class of attacks where an attacker modifies the
physical environment of the targeted chip in order to induce a fault during its
execution. The resulting failure can be used to extract sensitive information
(cryptographic keys) or bypass a PIN code verification for instance. More
generally, fault injection attacks give the ability to modify a program at runtime,
defeating static countermeasures that cannot foresee the failure (\textit{e.g.} secure
boot, access control).

The attacker's ability to subdue the system is highly dependant of its
experimental capacities: fault injection attacks can be performed with power glitches,
clock glitches, lasers, electromagnetic pulses, \textit{etc.} In this paper, we
propose an \gls{EMFI} attack against a \gls{SoC}. Electromagnetic pulses modify
the electric signals in the metallic chip wires. Faults are generated
when signals are modified during a small time window around the clock rising
edge. In this case, according to Ordas \textit{et al.}~\cite{OrdasGM15}, a
faulty value may be memorized.

Most previous hardware attacks target microcontrollers. Indeed, these chips are
slower and simpler than \glspl{SoC}. Therefore, an attacker can easily perturb
microcontrollers and exploit a fault.

\subsection{Motivation}
\label{sec:motivation}

The \gls{SoC} security model is focused almost entirely on software, so much
that hardware countermeasures such as TrustZone explicitly exclude hardware
security from their objectives. Fault injection attacks on \glspl{SoC} is a recent
research topic where some published articles
\cite{TimmersSW16,BADFET,MajericBB2016} focused on breaking software security
properties.

Fault injection is a versatile tool allowing to modify a program behaviour at
runtime, to inject vulnerabilities in a sound (and even proved)
software. The stake is the system security as a whole, as in controlling
what software is executed and what data can be accessed by whom. With software
security constantly improving and the costs and experimental difficulty of
performing fault injection attacks declining, we can surmise that the latter will become
a major threat in the future.

Understanding the \gls{EMFI} effects on a \gls{SoC} is required to understand
the threat and to design effective countermeasures. There is an extensive
literature on fault injection attacks on microcontrollers, and as a result, the most
secure devices against them are derived from microcontrollers (aka secure
components). The same work has to be done for \glspl{SoC}.

\subsection{Related work}
\label{sec:previous}

Since the seminal works of Boneh \textit{et al.}~\cite{boneh_faults} and Biham
and Shamir~\cite{biham_faults}, we know that fault injection attacks are a threat to the
security of cryptographic implementations. Researches on this topic has split in two
directions: on the one hand, the theoretical axis is interested in how a fault
can weaken the security of an application, mainly cryptographic algorithms
(\textit{e.g.}~\cite{piret_fa} and~\cite{giraud_fa} on AES).

On the other hand, the question is what faults can practically be performed in
systems. A formal description of these achievable faults is called a fault model
and it is this description that is used for theoretical analysis.

A fault model is always the interpretation of a physical behaviour at a
specific abstraction level. In other word, the same fault can be formalized
differently if we look at the transistor, the \gls{ISA} or the
software levels. Therefore, different studies looked at different levels for the
faults effects.

Several groups have developed fault models for
microcontrollers. Balasch \textit{et al.}~\cite{BalaschGV11} performed a clock
glitch attack on an 8-bit AVR microcontroller embedded in a smartcard. They show
that one can replace instructions in the execution flow by either targeting the
fetch or the execute stages of the 2-stage pipeline. 

ARMv7-M microcontrollers have been thoroughly studied, in particular the Cortex-M3
\cite{MoroDHRE13} and Cortex-M4 \cite{RiviereNRDBS15}. These papers highlight
that \gls{EMFI} on these devices disturbs the correct behaviour of the pipeline.
Moro \textit{et al.}~\cite{MoroDHRE13} show that a fault can modify the fetched
opcode (at the time of fault injection) and that the new opcode has no side
effects with high probability. It is therefore equivalent to a \texttt{NOP} instruction
(no-operation). Another observed effect is the modification of the data in a
\texttt{LOAD} instruction. According to Rivière \textit{et al.}~\cite{RiviereNRDBS15},
the fault suppresses the instructions fetch\footnote{up to 4 instructions may be fetched
in one clock cycle}. The previously loaded instructions are instead executed
again before resuming correct execution, and the disrupted instructions are
never executed. These works only focus on microcontrollers.

But they are simple systems, they have in-order pipelines, most of
the time only one core, a simple memory hierarchy (only L1
cache if any) and no support for virtual memory. More recent works have been
invested in trying to fault more complex processors, mainly ARM \glspl{SoC}.

Timmers \textit{et al.}~\cite{TimmersSW16} show how to attack an ARMv7-A chip by
taking control of the \gls{PC} with fault injection attacks. Cui and Housley~\cite{BADFET}
demonstrate an \gls{EMFI} targeting the communication between the \gls{CPU} and
the external memory chips. The authors insist particularly on the difficulty to
achieve the required temporal and spatial resolution for \gls{EMFI} on modern
\gls{SoC}. Majéric \textit{et al.}~\cite{MajericBB2016} discuss how to find the
correct \gls{EMFI} parameters and setup a fault injection on the AES
co-processor to a Cortex-A9 core.

A critical feature of modern \glspl{SoC} is their complex micro-architecture,
increasing the attack surface. As a consequence, it is now possible to obtain
hardware faults from software execution. Tang~\textit{et al.}~\cite{TangSS17}
achieve a fault by taking control of the microcontroller in charges of
monitoring and managing the energy of the \gls{SoC}. By software means, they are
able to modify the power voltage and the clock frequency. DRAM memories also are
susceptible to Rowhammer attacks~\cite{VeenFLGMVBRG16}: with specific access
patterns, one can switch bits into the memory chip. The gap between the
\gls{ISA} abstraction and its real implementation (notably in Out-of-Order
processors) leads to the Meltdown~\cite{MELTDOWN} and Spectre~\cite{SPECTRE}
vulnerabilities.

Today, we understand the risks due to the complex micro-architecture of modern
\glspl{SoC}. Fault injection attacks have been demonstrated to be a threat to these
systems. Proy \textit{et al.}~\cite{proy2019} have proposed a fault model characterization on \gls{SoC} (with an \gls{OS}) at the \gls{ISA} level. But the faults effects on the \gls{SoC}
micro-architecture have still not been evaluated.

Our work intends to bring this missing piece to our understanding of
the security model of modern \glspl{SoC}.

\subsection{Contributions}
\label{sec:contribution}

In this article, we focus on an ARMv8 \gls{SoC}, namely the Broadcom BCM2837 chip
at the heart of the \gls{RPi3}. It is a widely successful low cost single
board computer. This quad-core Cortex-A53 \gls{CPU} runs at
\SI{1.2}{\giga\hertz} and features a modern ``smartphone class'' processor. We
are using \gls{EMFI} and observe the resulting failures to deduce their origins. 

We observe radically new fault models that are neither described in other works
nor taken into account when discussing the modern embedded systems security.
In this paper, we demonstrate how we recovered these fault models and provide insights on the micro-architectural mechanisms leading to these models. 

The consequences are dire: a \gls{SoC} must not be considered as a black box
with respect to security. It is not enough to work on the software side security
if it does not rely on solid hardware foundations.\newline

The goal is to provide a micro-architectural explanation of the observed
behaviour. To that end, we have to control the targeted system and limit its
complexity. It implies most notably that we use a single-core configuration and
setup an identity mapping for virtual memory. This simplification choice does
not imply necessarily a harder exploitation on more realistic systems, since the discovered fault models would still be present.
But on such systems, it becomes hard to isolate the effect of a fault and attribute it to one subsystem: we cannot propose a simple model explaining the observed behaviour.

We describe our setup in
\autoref{sec:setup}, both the experimental apparatus used to inject faults but
also the targeted hardware and software environment.

To stress out the software layer impact on the observed failures, we
compare the faults observability with and without an \gls{OS} in section~\ref{sec:os}.

Observed faults on a bare-metal setup are analysed in sections \ref{sec:icache},
\ref{sec:mmu} and \ref{sec:l2cache}. For each fault category, we will explain
the process that allows us to infer the cause of the failure. The possibility to
exploit these faults will be discussed as well as the experimental difficulties
to achieve them. We finish with propositions to protect \glspl{SoC} against
these attacks in \autoref{sec:countermeasures} and conclude in
\autoref{sec:conclusion}.

\section{Fault injection on embedded systems}
\label{sec:fault_injection}
\subsection{The targeted chip: The BCM2837 on the Raspberry Pi 3 B}
\label{sec:setup}

\subsubsection{Presentation}

The \acrfull{RPi3} is a low cost single board computer. It features a
complete system able to run a complex OS such as Linux or Windows and their
applications. The \gls{SoC} powering this board is the BCM2837 from Broadcom, a
quad-core Cortex-A53 \gls{CPU} running at \SI{1.2}{\giga\hertz} with the help of
a dual core VideoCore IV GPU at \SI{400}{\mega\hertz}.

Our experiments are performed with our own software stack\footnote{Released as
  open-source software (MIT Licence). The git repository is available at
  \textit{blinded for reviews}}, with only one core active. To control the
behaviour of the chip, we have implemented the bare minimum to run our
applications: initialization of \gls{JTAG}, UART, GPIO, \gls{CPU} caches and
\gls{MMU}. We want to stress out that no \gls{OS} is running during our
experiment in sections \ref{sec:icache}, \ref{sec:mmu} and \ref{sec:l2cache}, to
avoid interference that could hinder our ability to infer the fault models. In
particular, we want to avoid the effects of context switching due to preemptive
scheduling by the \gls{OS}, the error recovery mechanisms (if an error occurs,
we want to know) and the caches maintenance performed by the \gls{OS}.

In order to later explain the causes of the failures observed, we describe
in more details two important subsystems of this \gls{SoC}: the cache hierarchy
and the \gls{MMU}.

\subsubsection{Cache hierarchy}

In modern systems, memory accesses are a lot slower than the \gls{ALU}. To avoid
loosing too much performance to this latency difference, small and fast memories
called caches are used to mirror a part of the memory space.

In the targeted \gls{SoC}, each core has two L1 caches (the smallest and fastest
kind), one dedicated to instructions (L1I), one dedicated to data (L1D). These
caches are \SI{16}{\kilo\byte} with \SI{64}{\byte} line width.

Then a second layer of cache, the L2 cache, is common to all cores and thus
provides a unified view of the memory space. Its size is \SI{512}{\kilo\byte}
with \SI{64}{\byte} line width.

\begin{figure}[h]
  \centering
    \includegraphics[width=0.6\textwidth]{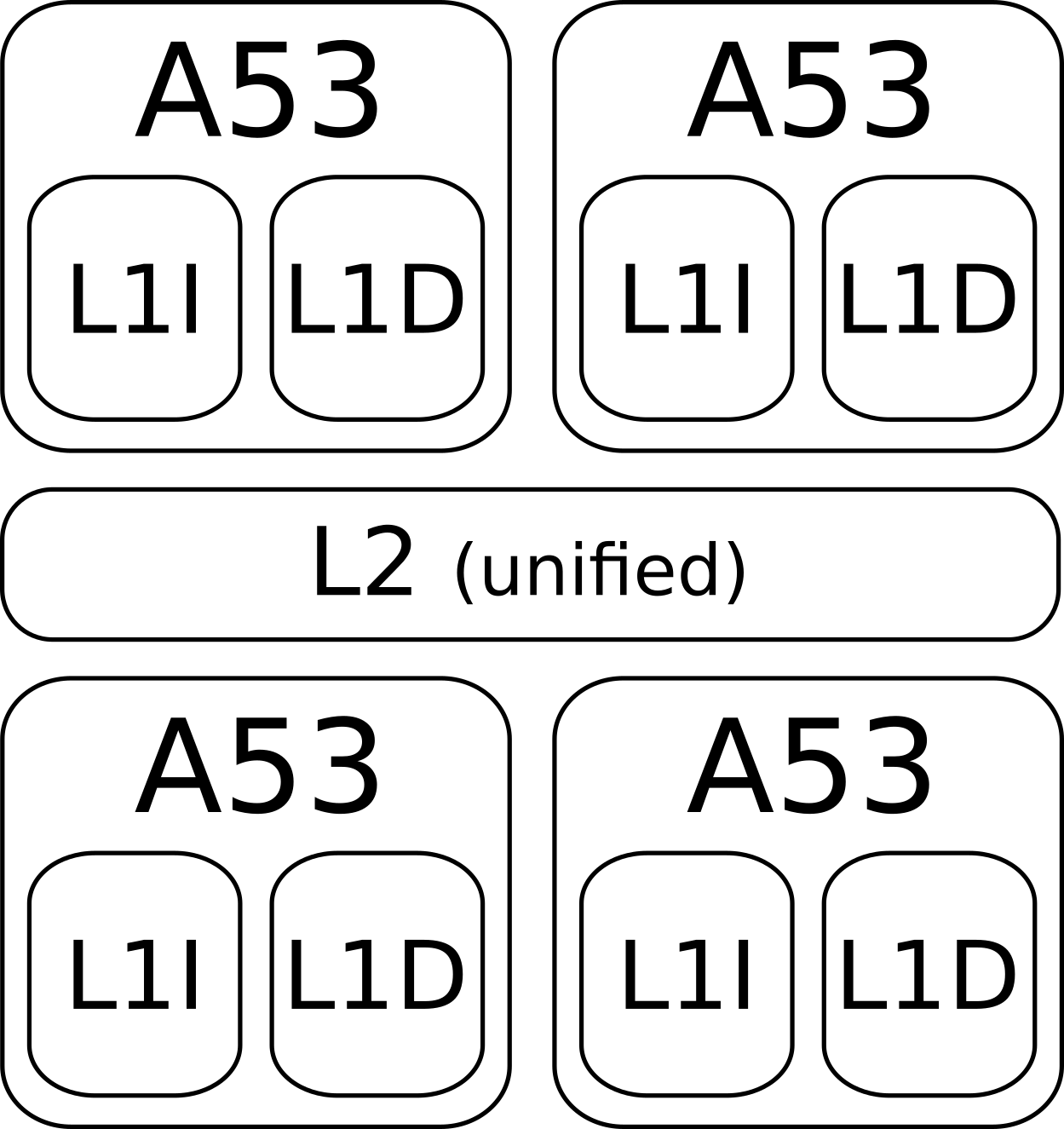}
    \caption{Memory hierarchy for the BCM2837.}
    \label{fig:memhier}
\end{figure}

\subsubsection{Memory management unit}

The \gls{MMU} is a central component for every multi-applications system. It
aims at virtualizing the physical memory of the system into a virtual one.
Therefore, the CPU only works with virtual addresses and during a memory access
to one of these addresses, the \gls{MMU} translates it into the corresponding
physical address which is transmitted to the memory controller of the system.
The information required for the translation of an address is called a \gls{PTE}
and it is stored in the physical memory and cached in the \gls{TLB}. There is a
\gls{PTE} for every allocated pages in the physical memory. Our bare metal
implementation allocates the whole address space with an identity mapping
(virtual and physical addresses are the same) with \SI{64}{\kilo\byte} pages.

In modern systems, the translation phase does not only compute the physical
address but also realizes different checks. These checks are monitoring if the
page can be written or not, which kind of process (user or supervisor) can
access it or should the page be stored in cache or not.

Among all its roles, the \gls{MMU} is also a security mechanism. Ensuring that
a read-only page cannot be written to and ensuring that only authorized processes
can access their corresponding pages. This last security mechanism is the memory
partitioning. On multi-applications systems, it avoids a process to spy or
corrupt the memory used by another process.

On complex \glspl{OS}, the \gls{MMU} and the \gls{PTE} are setup by the
kernel and are critical assets.


\subsection{The Electromagnetic fault injection bench}

To inject faults on the BCM2837, some apparatus is required. Our experimental
setup has been designed to be highly configurable and to work at higher
frequencies than most setups targeting microcontrollers. First, we use a
Keysight 33509B to control the delay between a trigger issued by the \gls{RPi3} board
before the instructions of interest. The Keysight 81160A generates the signal
for the EM pulse: one sinus period at \SI{275}{\mega\hertz} with a power of
\SI{-14}{\dBm}. A sinus is used instead of the usual pulse since it gives fewer
harmonics at the output of the signal generation chain. Then, this signal is
amplified with a Milmega 80RF1000-175 (\SI{80}{\mega\hertz} -
\SI{1}{\giga\hertz} bandwidth). Finally, the high power signal is connected to a
Langer RF U 5-2 near field probe. A part of this energy is therefore transmitted
into the metallic lines of the chip, which can lead to a fault.

The minimum latency between the initial trigger and the faulting signal reaching
the target is high: around \SI{700}{\nano\second}. As a consequence, the
targeted application must be long enough to be reachable by our fault injection
bench.

\subsection{Synchronization}

The main difficulty for fault injection is the synchronization: how to inject
precisely a fault on the targeted and vulnerable instructions.

To resolve this point, we need a temporal reference, given here by a GPIO: an
electrical rising edge is sent to a board pin by our application just before the
area of interest. Our setup is using the evaluator approach: the attacker can instrument the system to ease the experiments. In the case of a
real attack, the adversary would have to generate this trigger signal: it can be
done by monitoring communications, IOs, or EM radiation to detect patterns of
interest. In all cases, it is a tricky business highly application dependant.

But the trigger signal is just part of the problem: from this instant we must
wait the correct moment to inject the fault. To give a sense of the experimental
difficulty: for a chip running at \SI{1}{\giga\hertz}, a clock period lasts
\SI{1}{\nano\second}. In this lapse of time, light travels only for around
\SI{30}{\centi\meter}. Propagation times are not negligible. On a modern
\gls{SoC}, the matter is made more difficult by the memory hierarchy. Since
cache misses are highly unpredictable, they imply a corresponding jitter. It is
hard to precisely predict the duration of a memory access and therefore the time
to wait to inject the fault.

Synchronization is a problem, but not a hurdle that much. Indeed, the
attacker has only to inject faults until the correct effect is achieved. Because
of the jitter, for the same delay (time waited between fault injection and
trigger), different timing will be tested with respect to the running program.
If a fault with an interesting effect is possible, it will eventually be
achieved.

Additionally, as we will see in the next sections, memory transfers are
particularly vulnerable to \gls{EMFI}. They are also slower than the core
pipeline, allowing for a bigger fault injection timing window.

\subsection{How to change the fault effect?}

Fault effects are reproducible with a low ratio; meaning that if a fault has
been achieved, it will be achieved again with the same parameters but only for a
small ratio of the fault injections. In the other cases, no failure occurs or
another effect is observed (mostly due to jitter). To modify the fault effect,
the main parameters are the timing and the position of the probe over the
component. In particular, the signal parameters (shape, frequency, number of
periods) have an optimal value with respect to our requirements. The frequency
and the shape are chosen to maximize the EM coupling, the number of periods is
fixed to have the best timing precision.

\subsection{Forensic methodology}

As the targeted system is a closed box, we have a limited mean to
explore what is happening in the system, namely the \gls{JTAG}. With it, we are
able to halt the chip execution to read the register values and to read memory
as seen by a particular core (with a data viewpoint). Therefore, to pinpoint the
particular effects of a fault injection, we trade observability of the system
with controllability: we force the system state such that an observable change
gives us information on the fault mechanism. To maintain controllability, our
software footprint has to be minimal. As such we will not describe how to breach a
particular system with our faults since any exploit is highly application
dependant and our setup is not representative of a standard application
environment. Instead we will suggest exploit strategies: how such faults could
be used by a malicious attacker?

\newdate{date_raspbian}{18}{04}{2018}

\section{Impact of the operating system}
\label{sec:os}

To support our choice of bare metal applications to understand the fault model,
in this section we compare the faults observed with and without an \gls{OS} for
the same \gls{EMFI} parameters.

\subsection{Sensibility maps for the BCM2837}
Knowing where to place the probe for obtaining interesting effects is mandatory
for every perturbation experiments. Therefore, the first step consists in doing a
sensibility map of the component against EM perturbations.

During our experiments, two different setups were tested. The first one was
running the target program \autoref{lst:loop} on a bare metal system (one core
with only UART and \gls{JTAG} enabled). The second one was running the same
program as an application on a Linux-based OS\footnote{Raspbian Lite released on
  \displaydate{date_raspbian} available here:
  \url{https://downloads.raspberrypi.org/raspbian_lite/archive/2018-06-29-03:25/}}.

\begin{lstlisting}[language=Python,caption={Loop target
  application},label=lst:loop]
trigger_up();
//wait to compensate bench latency
wait_us(2);
invalidate_icache();
for(int i = 0; i<50; i++) {
  for(int j = 0; j<50; j++) {
    cnt++;
  }
}
trigger_down();
\end{lstlisting}

\begin{figure}
  \centering
  \includegraphics[width=.7\textwidth]{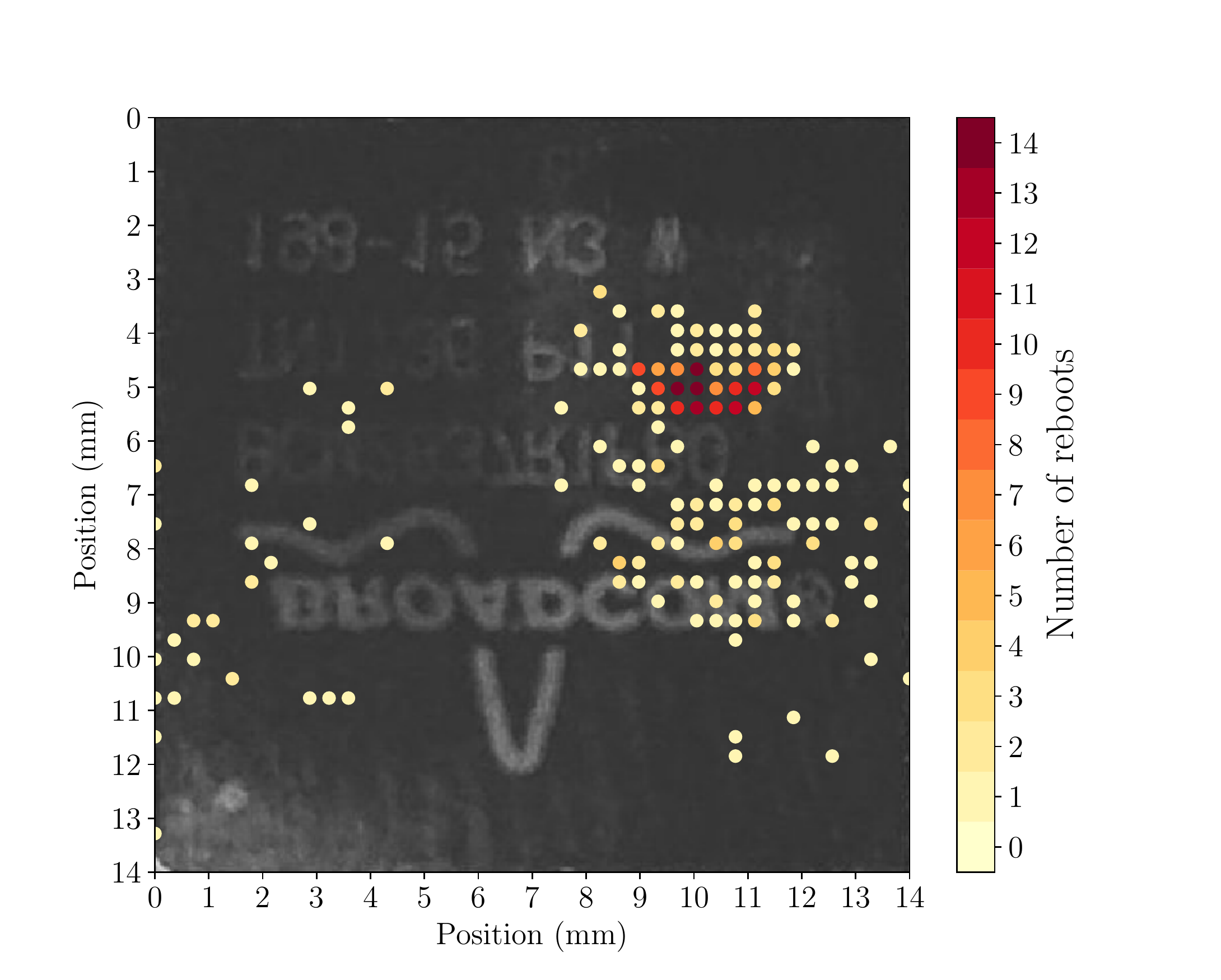}
  \caption{Bare metal sensitivity map}
  \label{fig:rpi3_carto_bare_metal}
\end{figure}

\begin{figure}
  \centering
  \includegraphics[width=.7\textwidth]{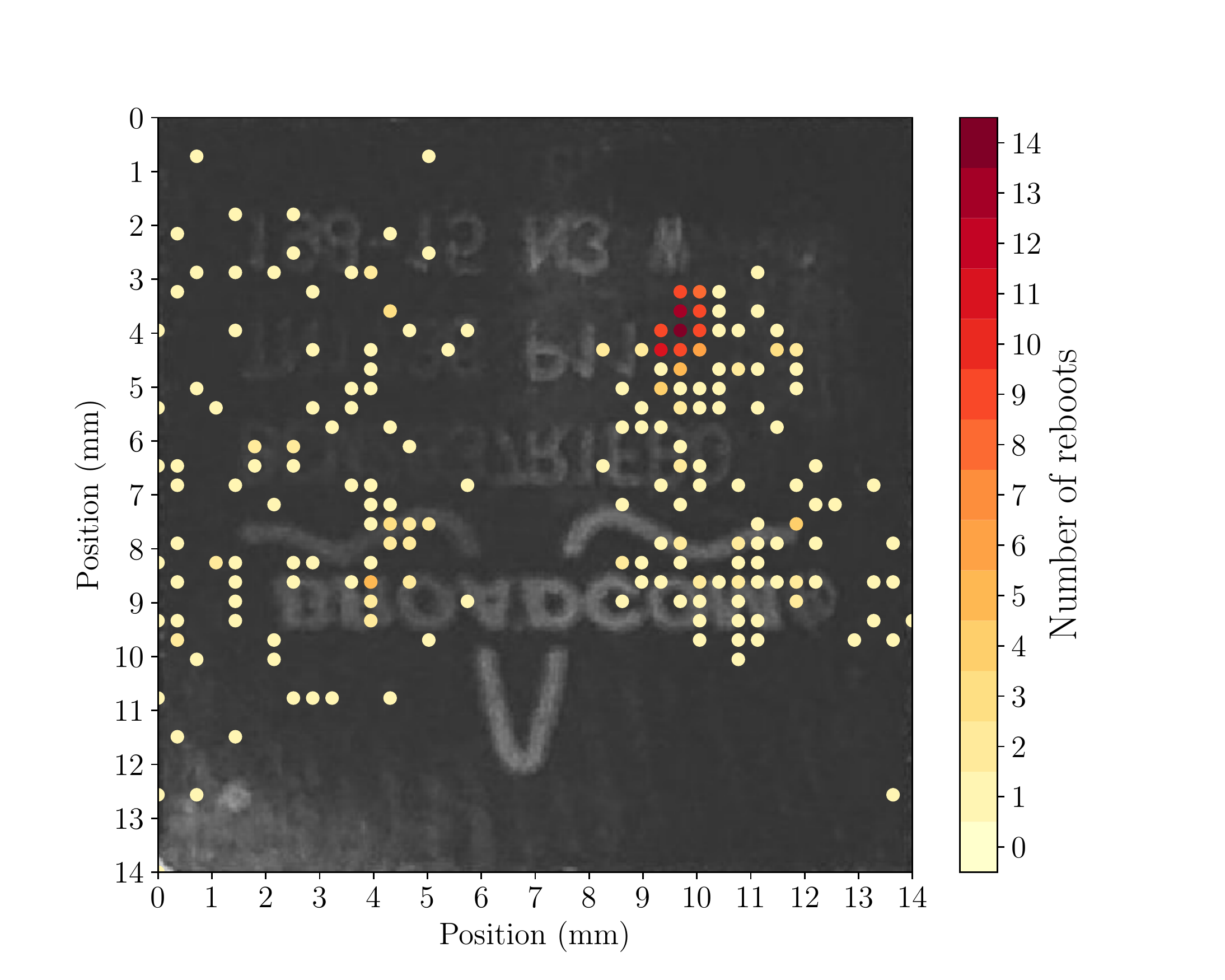}
  \caption{Linux sensitivity map}
  \label{fig:rpi3_carto_linux}
\end{figure}

The figures~\ref{fig:rpi3_carto_bare_metal} and~\ref{fig:rpi3_carto_linux} show the two sensibility maps with the number
of crashes induced by the perturbations for every probe location over the \gls{SoC} (for $27$ tries per location).
The area is divided in a $40$ per $40$ grid with a step of \SI{350}{\micro\meter}. This allows
us to cover the whole package of the \gls{SoC}. The first conclusion is that the
sensibility of the component under \gls{EMFI} depends on what is running
on it. The setup running with Linux has a wider sensitive area than the bare
metal one. However, the sensitive area of the bare metal setup is included in the
Linux one. This suggests that the two setups behave similarly under the
perturbations on this area. Since the Linux system embeds a far more complex
piece of software than the bare metal one, with more enabled interfaces, it may
explain that the Linux setup has a wider sensitive area.


\subsection{Faults on bare metal versus faults on Linux}

The sensitive areas are not the only differences of behaviour between the two setups.
Another difference is the impacts of the perturbations.
In other words, the faults obtained are different on the bare metal setup and the Linux setup.
More precisely, the observable effect of a fault as seen by the same application is different whether it runs on a bare metal setup or on a Linux setup.

Evaluating the fault model on the Linux setup is a complex
analysis.
With an \gls{OS}, faults are observable at the instruction level. This means the
effect induced by the perturbation is equivalent to a fault modifying one or
several instructions of the executed program.

On the bare metal setup, there are no observable faults at the instruction level
but instead at the micro-architectural level, as shown in the next sections. The
effect of the fault is equivalent to modifying the behaviour (signals or
configuration for instance) of subsystems like buses, \gls{MMU}, memory, caches,
etc.

This difference suggests that the usage of an \gls{OS} leads to a specific
perturbation behaviour. In this specific case, the fault model induced by the
\gls{OS} (instruction modification) is easier to exploit than the fault model on
bare metal setup: the \gls{OS} weakens the security of the system against fault
injection attacks.\newline

In this work, we will focus on the fault model on the bare metal setup. We will
show how to analyse and reconstruct the effect of a perturbation on the
micro-architectural elements of a processor.
 
However, we can suppose that the micro-architectural fault effect on the bare
metal setup explains the observed effect at the instruction level on the Linux
one.

\section{Fault on the instruction cache}
\label{sec:icache}

In this experiment, we achieve a fault in the L1 instruction cache (L1I).

\subsection{On the impossibility to fault the instruction execution flow}

Before reporting our positive results, we must report a negative one. Contrary
qto previous works on microcontrollers (where \textit{e.g.} instructions are
replaced with a \texttt{NOP} instruction), we are not able to prevent or modify directly
the execution of an instruction. In other words, in \autoref{lst:loop}, without
the \texttt{invalidate\_icache()} no faults are observed. Even if we cannot be
sure that no set of experimental parameters would achieve such a fault, we
thoroughly explored the parameters without success.

As we will see in this section and the following, all faults affect memory subsystems.
Probably because the buses involved in the memory transfers are easily coupled with our EM probe.

\subsection{The target}

The application targeted during this experiment is two nested loops shown on
\autoref{lst:loop}, executed after the L1I cache invalidation. No fault is
obtained without the invalidation. It is compiled without optimizations
(\verb|-O0|) since we do not want the compiler to optimize our code.

Since the instruction cache is invalidated before the loop execution, the
following instructions have to be (re)loaded in cache before their execution.
And it is this memory transfer that we will target with our fault injection. By
executing the same application with and without the cache invalidation and
measuring the duration of the high state of the trigger, we deduce that loading
instructions in cache has an overhead of \SI{2}{\micro\second}. Our bench has a
latency of \SI{700}{\nano\second}, so we can still hit this memory transfer. To
be able to observe the effect of a fault on the full timing range, a
\SI{2}{\micro\second} wait has been inserted between the trigger and the cache
invalidation.

\subsection{Forensic}

The fault is detected when the output sent back to the host is not $2500$ (the
value of \texttt{cnt} at the end of the program if everything went well). When a
fault is detected, we use the \gls{JTAG} to reexecute our application (in
\autoref{lst:loop}) by directly setting the \gls{PC} value at the start of the
loops. Then, we execute our program instruction by instruction while monitoring
the expected side effects (we do not inject faults anymore). All instructions are
correct except one, the counter incrementing instruction (the \texttt{add} instruction at
address \texttt{0x48a08} on \autoref{lst:loopasm}).

\begin{lstlisting}[language=C,caption={Loop target application assembly with -O0},label=lst:loopasm]
...
48a04:	b94017a0 	ldr	w0, [x29,#20]
48a08:	11000400 	add	w0, w0, #0x1
48a0c:	b90017a0 	str	w0, [x29,#20]
48a10:	b9401ba0 	ldr	w0, [x29,#24]
48a14:	11000400 	add	w0, w0, #0x1
48a18:	b9001ba0 	str	w0, [x29,#24]
48a1c:	b9401ba0 	ldr	w0, [x29,#24]
48a20:	7100c41f 	cmp	w0, #0x31
48a24:	54ffff0d 	b.le	48a04 <loop+0x48>
...
\end{lstlisting}

By monitoring the \texttt{w0} register before and after the \texttt{cnt}
increment instruction, we observe that the value is kept unchanged: the
increment is not executed. Since the fault is still present after the EM
injection, we conclude that a wrong instruction value is memorized in L1I. We
confirm this fault model by executing a L1I cache invalidation instruction
\texttt{ic iallu} (we set the \gls{PC} value to the instruction location in
memory). Reexecuting our application, the fault has disappeared.

We can infer that the injected fault has modified a value in the L1I cache, but
it is impossible to read the new value. Since the fault happens during the cache
filling, we can suppose that it is the memory transfer that has been faulted.

\subsection{Exploits}

This fault is one of the easiest to exploit since it is similar to the classical
instruction skip model. Therefore, most exploits based on this classical model
apply here. Since the faulted value is still present in the cache, it will stay
faulted until the cache is invalidated: we can call this model ``sticky
instruction skip''. Bukasa \textit{et al.}~\cite{Bukasa2018} demonstrate
applications of this fault model: hijacking the control flow and initiating a
\gls{ROP} attack among others.

\section{Targeting the MMU}
\label{sec:mmu}

The \gls{MMU} is a critical component of \glspl{SoC}. It is in charge of the
virtual memory mechanism. In this section, the fault changes the virtual to
physical memory mapping, albeit in an uncontrolled manner. The targeted
application is the same as in section~\ref{sec:icache}, shown on
\autoref{lst:loop}.



\subsection{The configuration of a working MMU}

To understand the effect of the fault, we begin to explore the state of a
working application (without any fault). This state is a legitimate one.

\subsubsection{Page tables}
The page tables are used to memorize the mapping between virtual and physical
memory. In our configuration, we have 3 \glspl{PTD} (mapping
\SI{512}{\mega\byte} chunks) and 
for each one, we have $8192$ \gls{PTE} pages of \SI{64}{\kilo\byte}. We show an
excerpt of the correct \glspl{PTE} on \autoref{fig:l3pt}.


\begin{figure}
  \centering
\begin{Verbatim}
0x00390000: 00000703 00000000 00010703 00000000
0x00390010: 00020703 00000000 00030703 00000000 
0x00390020: 00040703 00000000 00050703 00000000 
0x00390030: 00060703 00000000 00070703 00000000 
0x00390040: 00080703 00000000 00090703 00000000
\end{Verbatim}
    \caption{Memory dump excerpt for \glspl{PTE} before fault.\label{fig:l3pt}}
  \end{figure}

In the page tables, the most and least significant bits are used for the page
metadata (access rights, caches configuration, \textit{etc.}).

\subsubsection{TLB}

\glspl{TLB} (plural since there are several of them), are small buffers used to
speed up virtual to physical memory translation. As in a cache memory, the last
mappings are saved to be reused later without a full page tables walk by the
\gls{MMU}. In the targeted \gls{SoC}, \gls{TLB} hierarchy mirrors cache
hierarchy: the \gls{TLB} designates the unified Level2 buffer while
micro-\glspl{TLB} are dedicated to instructions or data in each core.

\subsubsection{Operating system}

In our bare metal application, all the pages are initialized in the page tables
with an identity mapping (virtual and physical addresses are identical). In a
system with an \gls{OS}, pages are allocated on-the-fly. On the one hand, this
simplifies the forensic analysis since we are sure that page tables are correct
prior to the fault. On the other hand, interesting faults may be missed if the
\gls{OS} page allocation is disrupted.

\subsection{Forensic}



To reconstruct the memory mapping, we use a pair of instruction computing the physical address (and the corresponding metadata) for a given virtual one.
A script has been designed to extract the memory mapping. By using the \gls{JTAG}, first the two instructions \verb|at s1e3r, x0; mrs x0, PAREL1| are written at a given address, then the
\texttt{x0} register is set to one virtual address, the two instructions are
executed and finally the \texttt{x0} register contains the corresponding
physical address.

With this method, we compare the memory mappings with (\autoref{fig:faulted_mapping}) and without (\autoref{fig:identity_mapping}) a fault.
 
\begin{figure}
    \centering
\begin{Verbatim}
VA      -> PA
0x0     -> 0x0      0x80000 -> 0x80000
0x10000 -> 0x10000  0x90000 -> 0x90000
0x20000 -> 0x20000  0xa0000 -> 0xa0000
0x30000 -> 0x30000  0xb0000 -> 0xb0000
0x40000 -> 0x40000  0xc0000 -> 0xc0000
0x50000 -> 0x50000  0xd0000 -> 0xd0000
0x60000 -> 0x60000  0xe0000 -> 0xe0000
0x70000 -> 0x70000  0xf0000 -> 0xf0000
\end{Verbatim}
    \caption{Correct identity mapping}
    \label{fig:identity_mapping}
\end{figure}

\begin{figure}
  \centering
\begin{Verbatim}
VA      -> PA
0x0     -> 0x0      0x80000 -> 0x0
0x10000 -> 0x10000  0x90000 -> 0x0
0x20000 -> 0x20000  0xa0000 -> 0x0
0x30000 -> 0x30000  0xb0000 -> 0x0
0x40000 -> 0x40000  0xc0000 -> 0x80000
0x50000 -> 0x50000  0xd0000 -> 0x90000
0x60000 -> 0x60000  0xe0000 -> 0xa0000
0x70000 -> 0x70000  0xf0000 -> 0xb0000
\end{Verbatim}
  \caption{Mapping after fault}
  \label{fig:faulted_mapping}
\end{figure}


Three different effects can be observed depending on the page:

\begin{itemize}
\item Pages are correct with an identity mapping up to \texttt{0x70000}.
  Remarkably theses are all the pages used to map our application in memory.
  Therefore, an hypothesis is that the corresponding translations are present in
  caches and are not impacted by the fault.
\item Pages are incorrectly mapped to \texttt{0x0}. A read at \texttt{0x80000}
  reads with success physical memory at \texttt{0x0}.
\item Pages are shifted. A read at \texttt{0xc0000} reads physical memory at
  \texttt{0x800000}.
\end{itemize}

If we invalidate the \gls{TLB} after a fault, nothing changes: the mapping stays
modified.
We conclude that the fault does not affect the cache mechanism of address translation  (at least what can be invalidated by software) but directly the \gls{MMU}.

To look for an explanation of the incorrect mapping,
we can look for the impact on page tables on \autoref{fig:l3xxxxpt}.

\begin{figure}
  \centering
\begin{Verbatim}
0x00390000: 40020607 00000000 40030607 00000000 
0x00390010: 40040607 00000000 40050607 00000000 
0x00390020: 40060607 00000000 40070607 00000000 
0x00390030: 40000000 8a210002 40000000 8a210002 
0x00390040: 400a0607 00000000 400b0607 00000000
\end{Verbatim}
  \caption{Memory dump excerpt for \glspl{PTE} after a fault.\label{fig:l3xxxxpt}}
\end{figure}

The fault on the \gls{MMU} has shifted the page tables in memory, and has
inserted errors in it. Since the memory translation is still valid after the
fault, and do not correspond to the shifted page tables, this shift is not the only source of incorrect translation.
Either the page walk is done from physical addresses and/or some \gls{TLB} are
not properly invalidated when we try to.

\subsection{Exploit}

This fault shows that the cornerstone of the key security feature in any
\gls{SoC}, namely memory isolation, does not withstand fault injection. In
\cite{drammer}, the authors use the rowhammer attack to fault a \gls{PTE}. The
faulted \gls{PTE} accesses the kernel memory which allow the attacker to obtain a privilege escalation: by overwriting an userland \gls{PTE} for accessing all the
memory,  by changing the user ID to root or by changing the entry point of an executable.

Additionally, this fault model is a threat to pointer authentication
countermeasures, as proposed in the recent ARMv8.3 ISA. This pointer protection
works by storing authentication metadata in the most significant bits (usually
useless) of a pointer value. To use a pointer, the chip first validates the
authentication metadata. In our case, the attacker does not need to alter the
pointer value, it can alter where it physically points to, at a coarse (page)
granularity.

\section{Shifting data chunks in L2}
\label{sec:l2cache}

Another interesting behaviour when faulting a modified version of the loop
target, \textit{cf} listing~\ref{lst:register_transfer}, was investigated with
\gls{JTAG}.

\begin{lstlisting}[language=C,caption={Targeted assembly},label=lst:register_transfer]
489bc:	a9be7bfd 	stp	x29, x30, [sp,#-32]!
489c0:	910003fd 	mov	x29, sp
489c4:	b9001fbf 	str	wzr, [x29,#28]
489c8:	b9001bbf 	str	wzr, [x29,#24]
489cc:	b90017bf 	str	wzr, [x29,#20]
489d0:	900001a0 	adrp	x0, 7c000 <RT2+0xb0>
489d4:	912d2000 	add	x0, x0, #0xb48
489d8:	d2802002 	mov	x2, #0x100
489dc:	52800001 	mov	w1, #0x0
489e0:	94000b28 	bl	4b680 <memset>
489e4:	97fefe67 	bl	8380 <fast_trig_up>
489e8:	d2800040 	mov	x0, #0x2
489ec:	97feffe2 	bl	8974 <wait_us>
489f0:	94008765 	bl	6a784 <invalidate_icache>
489f4:	940087ad 	bl	6a8a8 <init_registers>
489f8:	b9001fbf 	str	wzr, [x29,#28]
489fc:	14000010 	b	48a3c <loop+0x80>
48a00:	b9001bbf 	str	wzr, [x29,#24]
48a04:	14000008 	b	48a24 <loop+0x68>
48a08:	940087c1 	bl	6a90c <register_transfer>
48a0c:	b94017a0 	ldr	w0, [x29,#20]
48a10:	11000400 	add	w0, w0, #0x1
48a14:	b90017a0 	str	w0, [x29,#20]
48a18:	b9401ba0 	ldr	w0, [x29,#24]
48a1c:	11000400 	add	w0, w0, #0x1
48a20:	b9001ba0 	str	w0, [x29,#24]
48a24:	b9401ba0 	ldr	w0, [x29,#24]
48a28:	7100c41f 	cmp	w0, #0x31
48a2c:	54fffeed 	b.le	48a08 <loop+0x4c>
\end{lstlisting}

We observe that data are shifted in the L2 cache, as if addresses had been slightly modified in the memory transfer writing to L2.

\subsection{Forensic}

A quick step by step execution shows that we are trapped into an
infinite loop. A \gls{JTAG} memory dump at the instruction memory location shows
modified instructions as seen on \autoref{dump:L2unified}, to be compared to the unfaulted dump on \autoref{dump:L2ok}.

With the same parameters, several similar faults have been obtained with two
observed faulty memory dumps, here called F1 and F2. Similar because the same
infinite loop is obtained (as shown in step by step execution), but the memory
dumps obtained from \gls{JTAG} are slightly different. We will show that the
difference is due to discrepancies between caches for F2 and that invalidating
L1I cache restore coherence with the F1 results.

\begin{figure}[h]
    \centering
\begin{Verbatim}
0x000489d8: d2800040 97feffe2 00000002 00000008 
0x000489e8: 00000002 00000008 \textcolor{red}{910003fd b9001fbf}
0x000489f8: \underline{\textcolor{red}{b9001bbf b90017bf} \textcolor{blue}{11000400 b90017a0}}
0x00048a08: \underline{\textcolor{blue}{b9401ba0 11000400} \textcolor{darkgreen}{b9001ba0 b9401ba0}}
0x00048a18: \underline{\textcolor{darkgreen}{7100c41f 54fffeed}} b9401fa0 11000400 
0x00048a28: b9001fa0 b9401fa0 81040814 77777777 
\end{Verbatim}
    \caption{Memory dump showing the instructions in the infinite loop as seen
      by the \gls{JTAG} for F1. The instructions in the infinite loop are
      underlined. \label{dump:L2unified}}
\end{figure}

 
\begin{figure}[h]
    \centering
\begin{Verbatim}
0x000489b8: d65f03c0 a9be7bfd \textcolor{red}{910003fd b9001fbf}
0x000489c8: \textcolor{red}{b9001bbf b90017bf} 900001a0 912d2000
0x000489d8: d2802002 52800001 94000b28 97fefe67 
0x000489e8: d2800040 97feffe2 94008765 940087ad
0x000489f8: b9001fbf 14000010 \textcolor{brown}{b9001bbf 14000008} 
0x00048a08: \textcolor{brown}{940087c1 b94017a0} \textcolor{blue}{11000400 b90017a0} 
0x00048a18: \textcolor{blue}{b9401ba0 11000400} \textcolor{darkgreen}{b9001ba0 b9401ba0} 
0x00048a28: \textcolor{darkgreen}{7100c41f 54fffeed} b9401fa0 11000400
\end{Verbatim}
    \caption{Memory dump showing correct (without fault) instructions for the same memory region.\label{dump:L2ok}}
\end{figure}

 
The step by step execution is coherent with the instructions shown by the JTAG
dump for F1 (registers are loaded, stored and incremented as specified by the
instructions in this dump, not the correct ones). Since \gls{JTAG} dump and
instructions execution are coherent, it seems that the fault is present in L2
cache (unified view). The reconstructed instructions can be seen in listing \ref{lst:l2faulted}.

\begin{lstlisting}[language=C,caption={Assembly reconstruction of the faulted instructions},label=lst:l2faulted]
489f8:	b9001bbf 	str	wzr, [x29,#24]
489fc:	b90017bf 	str	wzr, [x29,#20]
48a00:	11000400 	add	w0, w0, #0x1
48a04:	b90017a0 	str	w0, [x29,#20]
48a08:	b9401ba0 	ldr	w0, [x29,#24]
48a0c:	11000400 	add	w0, w0, #0x1
48a10:	b9001ba0 	str	w0, [x29,#24]
48a14:	b9401ba0 	ldr	w0, [x29,#24]
48a18:	7100c41f 	cmp	w0, #0x31
48a1c:	54fffeed 	b.le	489f8
\end{lstlisting}
 
\paragraph{MMU} After verification as in section~\ref{sec:mmu}, we observe that
the \gls{MMU} mapping is indeed modified but not for the memory region of
interest. In particular, memory addresses from \texttt{0x0} to \texttt{0x7FFFF}
are still correctly mapping identically virtual and physical addresses. The
fault on the \gls{MMU} cannot therefore explain our observations.

\paragraph{L1 fault} The L1I is of course modified by the fault since the
behaviour of the application is modified. But the execution is coherent with the
\gls{JTAG} dump (showing the L1D cache) hinting that L2 is impacted too. As seen
on figure~\ref{dump:L2nonunified}, the dump
after F2 shows different values in L1D memory with respect to F1 (for the first
two instructions), but the execution trace is coherent with F1 \gls{JTAG} dump
(on figure~\ref{dump:L2unified}), not with F2 \gls{JTAG} dump (on
figure~\ref{dump:L2nonunified}).

\begin{figure}[h]
    \centering
\begin{Verbatim}
0x000489f8: \underline{\textcolor{brown}{940087c1 b94017a0} \textcolor{blue}{11000400 b90017a0}} 
0x00048a08: \underline{\textcolor{blue}{b9401ba0 11000400} \textcolor{darkgreen}{b9001ba0 b9401ba0}} 
0x00048a18: \underline{\textcolor{darkgreen}{7100c41f 54fffeed}} b9401fa0 11000400 
0x00048a28: b9001fa0 b9401fa0 7100c41f 54fffded
\end{Verbatim}
    \caption{Memory dump showing the instructions in the infinite loop as seen
      by the \gls{JTAG} for F2.\label{dump:L2nonunified}}
\end{figure}

As it turns out, \texttt{940087c1} encodes a branching instruction that is not
followed in the step by step execution. In the case of F2, the \gls{JTAG} dump
does not reflect the values in the L1I cache. If we invalidate L1I cache,
nothing change (either in the execution trace or in the \gls{JTAG} dump). But if
we invalidate to point of coherency at address \texttt{0x489f8}, then the new \gls{JTAG}
dump becomes the same as for F1.

\paragraph{Hypotheses} The effect of these faults seem to manifest in the L2
cache. We can observe that it consists in shifting groups of 4 instructions (128
bits or \SI{16}{\byte}, under the cache line size of \SI{64}{\byte}) at a nearby
memory location. 128-bit is the size of the external memory bus connected to L2. A fault model could be that the address corresponding to a cache transfer
toward L2 has been modified by a few bits.

\begin{figure}[h]
  \centering
    \includegraphics[width=0.8\textwidth]{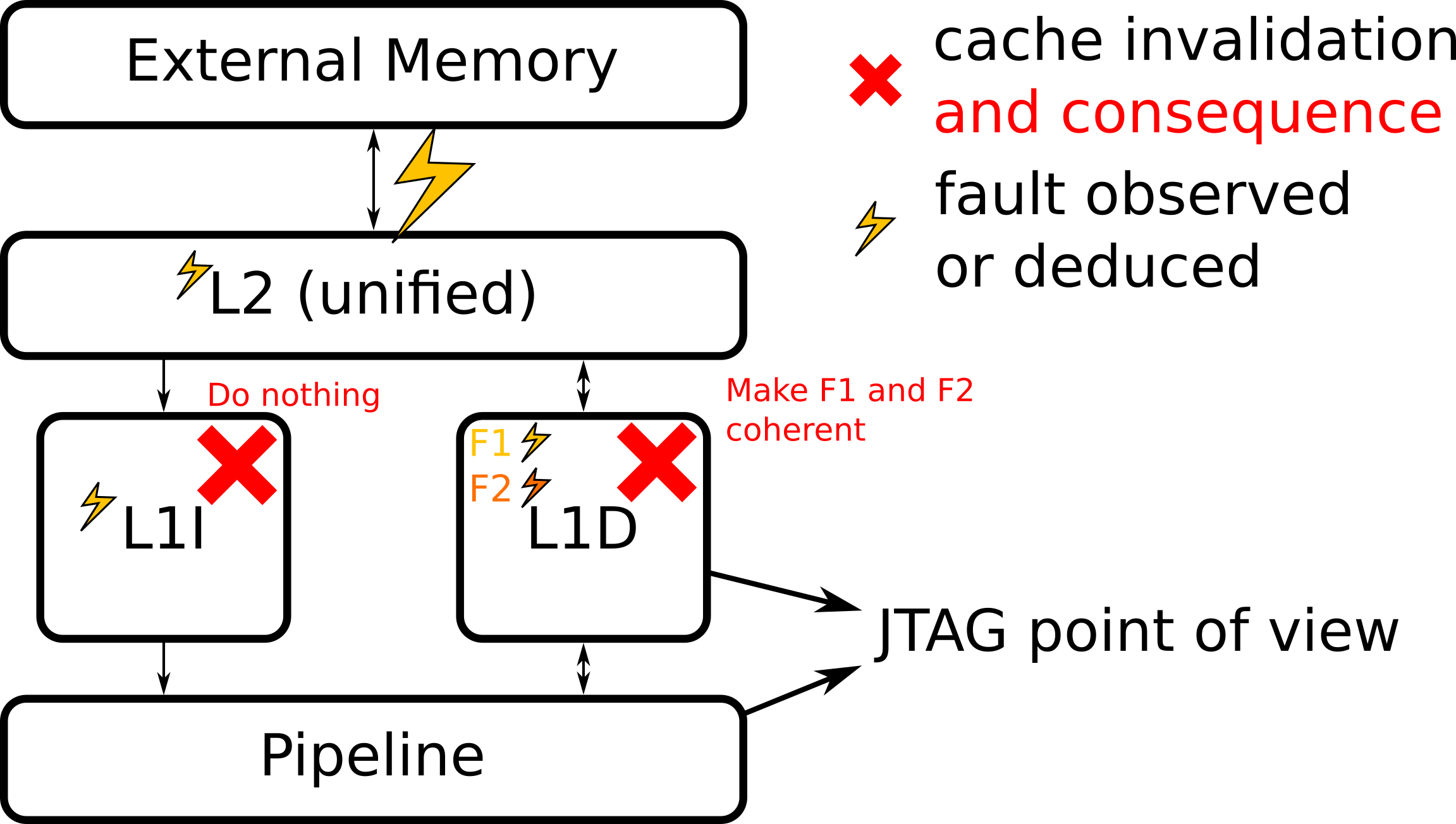}
    \caption{Summary of our fault model for the faults observed in \autoref{sec:l2cache}.}
    \label{fig:l2sum}
\end{figure}

The fault model is summarized on \autoref{fig:l2sum}. But our model has some limits: the presence of F2 means that, with one fault, values in L2 and in L1D are modified simultaneously in an uncoherent way.
Is it due to the EM probe coupling with several buses or to a micro-architectural mechanism ?

\subsection{Exploits}

A fault in the L2 cache can impact either the instructions or the data giving
more power to the attacker. Yet she does not control in what way the memory
will be modified. Why these particular 16 bytes blocks have been shifted?
Nevertheless, just corrupting data or instruction randomly is often enough to
achieve the desired effect (as in the \texttt{NOP} fault model).

\section{Countermeasures}
\label{sec:countermeasures}


The vulnerabilities described above can be summarized as integrity and authenticity problems: instructions
(or data) are altered between their storage in memory and their reuse. 
If we consider that only a legitimate entity can write to memory, then integrity is the only problem.
In the other case, we must also ensure the write authenticity.

\subsection{Ensuring integrity}

Ensuring the integrity of signals in a chip is a well known problem that had to be solved to allow the use of chips in harsh environments: space or nuclear reactors for example.
Yet one must account for a different threat model: when designing for security, the fault value must be considered intentional, not drawn from a uniformly random distribution.

To ensure integrity, designers use redundancy. We can duplicate the core, executing exactly the same things on both cores and verifying that results are identical. Redundancy can also be achieved using error detection codes. Of course, redundancy has an overhead but this is the cost of guaranteeing the integrity.

When dealing with the threat of physical attacks, ensuring integrity is often not enough. If we consider that the attacker can modify memory, she could bypass the error detection code or write the same error value in the duplicated memories.
In this case, we must ensure the data authenticity.
With cryptography, authenticity can be guaranteed by relying on \glspl{MAC}.
We can imagine a strategy based on this principle to be able, in hardware, to detect changes in data or instructions.

\subsection{MAC Generation}

A solution, to ensure authenticity has been proposed with SOFIA~\cite{SOFIA}.
For each data or instruction block (the block size has to be adapted to the
micro-architecture), the objective is to calculate and associate a \gls{MAC} to
detect any alteration. In addition to the data itself, the \gls{MAC} calculation
must also be address dependent to detect shifts between an address and the
corresponding data (as observed in the L2 cache in section \ref{sec:l2cache}).
\glspl{MAC} must be generated at the right time: in the case of instructions,
they can be computed at compile time. But for the data, it must be possible to
do the generation at the pipeline output (during memory access).

\subsection{MAC Verification}
Depending on the energy consumption/performance trade-off, two implementation strategies can then be considered on a
system-wide basis to perform the \gls{MAC} check.

\begin{itemize}
\item \textbf{Just-in-time:} To minimize the overall activity, this
  strategy consists in bringing the data and its \gls{MAC} back to the \gls{CPU}
  (or \gls{MMU}) in a classical way. The verification would then be performed by
  the \gls{CPU} just before data consumption. In case of mismatch, a request has
  to be made to the higher memory level (the L1 cache) to perform a
  verification on its own data version. If the new verification is
  successful, the data would be transmitted to the \gls{CPU} and if not, a
  request to the next level (L2) has to be made. Thus, several checks are
  performed only when necessary, but the cost of a mismatch (that can be due to
  environmental radiations) is very high.
\item \textbf{Proactive:} A second strategy can be used to reduce the time
  penalty in case of error. Integrity checks are automatically performed at each
  level of the memory hierarchy. Thus, errors can be detected early, before they
  reach the \gls{CPU}. Each cache level can ensure independently that it has
  only valid data, but the energy consumption would be higher.
\end{itemize}

\section{Conclusion}
\label{sec:conclusion}

In this paper, we have demonstrated some vulnerabilities with respect to fault injection
attacks specific to \gls{SoC}. In particular, the memory hierarchy and the
\gls{MMU} can be altered which creates a mismatch between how the hardware
behave and what the software expect. Nowadays, the computing systems security
is focused on the software side, there are no efficient countermeasure against
hardware perturbations in modern \gls{SoC}. Pointer
authentication as proposed by the ARMv8.3 \gls{ISA}, for example, does not
resist the introduced fault model.

Exploitation depends heavily on the interaction between the hardware (the
specific device) and the software (including application and \gls{OS}). Therefore
the cautious developer cannot predict where vulnerabilities will occurs and as a
consequence cannot efficiently protect its application. Today, such attacks
using \gls{EMFI} are still quite hard to realize: they require expensive
apparatus, human resources to do the experiments, etc. But they are within reach
of small organizations and we can expect that the difficulty and cost of these
attacks will be lower in the future.

Actions must be taken to ensure that computing systems handle
sensitive information securely. The performances/energy consumption trade-off has been
settled by implementing the two kinds of cores in the same \gls{SoC}. In the
same way, we need \textit{secure} cores as well that compromise both on
performances and energy consumption but can offer much stronger security
guarantees.

\bibliographystyle{alpha}
\bibliography{biblio} 

\end{document}